\documentclass[aps,prb,twocolumn,superscriptaddress,oneside,floatfix,amsmath,showpacs,amssymb,byrevtex]{revtex4-1}
\usepackage{graphicx}
\usepackage{dcolumn}
\usepackage{bm}
\usepackage[usenames]{color}
\usepackage{hyperref}

\begin{document}

\title{Magneto-optical Kramers-Kronig analysis}

\author{Julien Levallois, Ievgeniia Nedoliuk, Iris Crassee, Alexey B. Kuzmenko}
\affiliation{Department of Quantum Matter Physics, University of Geneva, CH-1211 Geneva 4, Switzerland}
\email{Alexey.Kuzmenko@unige.ch}

%\author{Julien Levallois}
%\affiliation{Department of Quantum Matter Physics, University of Geneva, CH-1211 Geneva 4, Switzerland}
%
%\author{Ievgeniia Nedoliuk}
%\affiliation{Department of Quantum Matter Physics, University of Geneva, CH-1211 Geneva 4, Switzerland}
%
%\author{Iris Crassee}
%\affiliation{Department of Quantum Matter Physics, University of Geneva, CH-1211 Geneva 4, Switzerland}
%
%\author{Alexey B. Kuzmenko}
%\affiliation{Department of Quantum Matter Physics, University of Geneva, CH-1211 Geneva 4, Switzerland}
%\email{Alexey.Kuzmenko@unige.ch}

\date{\today}

%%%%%%%%%%%%%%%%%%%%%%%%%%%% ABSTRACT

\begin{abstract}
We describe a simple magneto-optical experiment and introduce a magneto-optical Kramers-Kronig analysis (MOKKA) that together allow extracting the complex dielectric function for left- and right-handed circular polarizations in a broad range of frequencies without actually generating circularly polarized light. The experiment consists of measuring reflectivity and Kerr rotation, or alternatively transmission and Faraday rotation, at normal incidence using only standard broadband polarizers without retarders or quarter-wave plates. In a common case, where the magneto-optical rotation is small (below $\sim$ 0.2 rad), a fast measurement protocol can be realized, where the polarizers are fixed at 45$^\circ$ with respect to each other. Apart from the time-effectiveness, the advantage of this protocol is that it can be implemented at ultra-high magnetic fields and in other situations, where an \emph{in-situ}  polarizer rotation is difficult. Overall, the proposed technique can be regarded as a magneto-optical generalization of the conventional Kramers-Kronig analysis of reflectivity on bulk samples and the Kramers-Kronig constrained variational analysis of more complex types of spectral data. We demonstrate the application of this method to the textbook semimetals bismuth and graphite and also use it to obtain handedness-resolved magneto-absorption spectra of graphene on SiC.
\end{abstract}


\maketitle

%%%%%%%%%%%%%%%%%%%%%%%%%%%% INTRODUCTION

\section{Introduction}

Applying a magnetic field to materials influences profoundly their optical properties by the Zeeman splitting, formation of Landau levels (LLs), modification of the optical selection rules and sometimes inducing phase transitions. The strength of magneto-optical (MO) effects depends on many factors, such as the density and mobility of the charge carriers, the value of magnetic moments and their interaction, the band structure and the spin-orbit coupling. Magnetic field breaks the time reversal symmetry (unless it is already broken spontaneously) and triggers non-reciprocal optical phenomena, such as the Faraday \cite{FaradayPTRS1846} and the Kerr \cite{KerrPM1877} rotation. The MO rotation, which was historically the first compelling evidence of the electromagnetic nature of light, plays nowadays a crucial role in telecommunications, data storage, laser technology and material characterization.

It is well-known that magneto-resistivity and Hall-effect measurements are in general more informative than the DC zero-field transport. Likewise, the magneto-optical response carries more ample information about a physical system than zero-field optics. It is also more complicated from the experimental and theoretical point of view. While the linear electromagnetic response of an isotropic material without field is described by a scalar function, such as the complex optical conductivity $\sigma(\omega)$ or the dielectric function $\epsilon(\omega) = 1 + 4\pi i\sigma(\omega)/\omega$, its MO counterpart is essentially a $3\times 3$ tensor. The Onsager reciprocity relations \cite{OnsagerPR31} impose on it the following structure:
\begin{equation}\label{condtensor3D}
\hat{\epsilon}(\omega,B)=\left[
\begin{matrix}
\epsilon_{xx}(\omega,B) & \epsilon_{xy}(\omega,B) & 0\\
-\epsilon_{xy}(\omega,B) & \epsilon_{xx}(\omega,B) & 0\\
0 & 0 & \epsilon_{zz}(\omega,B)\\
\end{matrix}\right]
\end{equation}
\noindent which involves three complex functions, $\epsilon_{xx}$, $\epsilon_{xy}$ and $\epsilon_{zz}$, instead of one (the field $B$ is along the $z$ axis). If the light polarization is confined in the $xy$ plane, which is the case for normal-incidence experiments in the Faraday geometry, where the field is parallel to the propagation of light), then only the two components, $\epsilon_{xx}$ and $\epsilon_{xy}$, play a role. This allows us to restrict our discussion to the $2\times 2$ tensor (we omit the $B$ argument for brevity):
\begin{equation}\label{condtensor}
\hat{\epsilon}(\omega)=\left[
\begin{matrix}
\epsilon_{xx}(\omega) & \epsilon_{xy}(\omega) \\
-\epsilon_{xy}(\omega) & \epsilon_{xx}(\omega)
\end{matrix}\right],
\end{equation}
\noindent which has the eigenvalues $\epsilon_{\pm}(\omega)= \epsilon_{xx}(\omega)\pm i
\epsilon_{xy}(\omega)$ corresponding to the right- and the left-handed circular polarizations. It is notable that this reduction is valid not only for fully isotropic but also for uniaxial media with the optical axis along the $z$ direction, which increases dramatically the number of materials to which our discussion applies.

The goal of a complete linear-response optical-spectroscopy experiment in zero field is to determine the real and the imaginary parts of the dielectric function. While certain techniques, such as spectroscopic ellipsometry or time-domain spectroscopy, can provide experimental access to both of them, other methods rely on the Kramers-Kronig (KK) relations\cite{KramersNature26,KronigJOSA26,LandauLifshitz,DresselGruner} due to a limited set of measured optical quantities. A widespread technique of this kind is reflectance spectroscopy, where one measures the absolute reflectivity $R(\omega)$ in a broad range of frequencies and determines the phase $\theta(\omega)$ of the complex reflectivity coefficient\cite{Note1}:
\begin{equation}\label{r}
r(\omega) = \sqrt{R(\omega)}e^{i\theta(\omega)}=\frac{1-\sqrt{\epsilon(\omega)}}{1+\sqrt{\epsilon(\omega)}}.
\end{equation}
\noindent Jahoda found\cite{JahodaPR57} that the KK relation can be applied to the function $f(\omega)=\ln r(\omega)=\ln \sqrt{R(\omega)} + i\theta(\omega)$:
\begin{equation}\label{kkrinf}
\theta(\omega) = -\frac{1}{\pi}\wp\int_{-\infty}^{\infty}\frac{\ln \sqrt{R(x)}}{x-\omega}dx + \pi,
\end{equation} which can be used to restore the unknown complex phase by a direct integration of the measured reflectivity properly extrapolated outside the experimental range. Due to the parity relation $R(-\omega) = R(\omega)$ this formula reduces to
\begin{equation}\label{kkr}
\theta(\omega) = -\frac{\omega}{\pi}\wp\int_{0}^{\infty}\frac{\ln R(x)}{x^2-\omega^2}dx + \pi.
\end{equation}

\noindent The knowledge of the phase together with the reflectivity itself allows, via the Fresnel equations or other boundary-conditions, the determination of other optical functions, including the dielectric function and the optical conductivity.

In magneto-optics, a complete experiment should allow determination of the real and the imaginary parts for both $\epsilon_{xx}(\omega)$ and $\epsilon_{xy}(\omega)$ or equivalently for both $\epsilon_{\pm}(\omega)= 1 + 4\pi i\sigma_{\pm}(\omega)/\omega$. Obviously, the number of quantities to be measured should be doubled as compared to the case of $B=0$. At a first glance, one should simply work with circular polarized light and apply the zero-field techniques to each handedness separately. Although conceptually appealing, this strategy is difficult to realize in broadband spectroscopy as it is experimentally challenging to generate circular polarized light in a broad range of frequencies. Standard waveplates, typically made of quartz, operate at selected wavelengths. There are reports of multilayer waveplates that cover a continuous range of frequencies, which is however still somewhat limited \cite{MassonOL06,KaveevAO13}. Many waveplates must therefore be combined to cover a reasonably broad spectrum of circular polarized light, which is especially hard in the range of the phonon absorption in quartz. An alternative solution using elliptically polarized light was also demonstrated \cite{KarraiPRL92} that allows reducing the number of waveplates but significantly complicates the analysis. A broadband phase-shifting element (retarder) using an internal reflection in silicon and other materials was fabricated \cite{KangTSF11,XiIPT14}. Working with it is however associated with extra light absorption, the need of a delicate alignment and the penalty (in the case of Si) of not seeing visible light after the retarder.

An alternative broadband solution without generating circular light is based on polarimetry, where one can measure the MO rotation and ellipticity, or related quantities. A rather sensitive polarimetric method is based on a photoelastic modulation (PEM), where one can  measure these quantities using a lock-in demodulation at the main PEM frequency and at its second harmonic \cite{SatoJJAP81,YamaguchiPRB98}. However, in order to achieve an experiment that is complete in the sense indicated above, one has to measure additionally the complex reflectivity or transmission. Furthermore, the calibration procedure for the PEM-based measurement is quite involved. Another relevant technique is a spectroscopic magneto-ellipsometry with an access to many Muller-matrix elements \cite{KuhneRSI14}. While allowing a number of independent quantities to be measured separately, this experiment unavoidably mixes the $xy$ and $z$ components of the $3\times 3$ MO tensor due to the finite angle of incidence and thus adds extra complexity to the analysis. To our knowledge, a model-independent extraction of the complex functions $\epsilon_{+}(\omega)$ and $\epsilon_{-}(\omega)$ by this method has not been demonstrated yet.

In this paper we propose a rather simple and fast pathway towards a complete broadband magneto-optical spectroscopy of materials with the MO conductivity described by Equation (\ref{condtensor}). It involves the measurement of the reflectivity (or transmission) and the Kerr (or Faraday) rotation at a normal, or near-normal angle of incidence using two conventional polarizers without the need to measure other quantities, such as the ellipticity. The treatment of this set of data relies on a generalized magneto-optical KK analysis (MOKKA), which we introduce and mathematically justify in this paper.

To emphasize the basic ideas of the method, we shall restrict our consideration to materials, where the magnetic and magneto-electric susceptibility are negligible as compared to the dielectric function, which is valid in a vast majority of non-magnetic compounds. In the discussion, we shall however speculate on a possible inclusion of these electromagnetic terms that play an essential role in magnetic materials \cite{SieversPR63}, multiferroics\cite{PimenovNP06} and topological insulators \cite{QiPRB08}. For the sake of simplicity we shall also assume that the Fresnel equations are satisfied, even though the essential part of MOKKA holds even in the non-local limit, where more general electromagnetic relations should be used.

The rest of the paper is organized as follows. Section \ref{SectionTheory} is devoted to the generalization of the conventional Kramers-Kronig reflectivity formalism to the case, where the magnetic field is set perpendicular to the sample surface. In section \ref{SectionVariational}, the variational KK-constrained analysis\cite{KuzmenkoRSI05} is extended to MO experiments. Section \ref{SectionExp} describes the experimental technique. Three application examples are presented in section \ref{SectionApp}. The results are summarized in Section \ref{SectionSummary}.

\section{Direct Kramers-Kronig analysis of magneto-reflectivity and Kerr rotation}
\label{SectionTheory}

We start by considering the magneto-reflectivity and Kerr rotation spectra of a bulk sample at normal incidence. The complex magneto-reflectivity tensor has the same symmetry as the MO tensor (\ref{condtensor}):
\begin{equation}\label{rtensor}
\hat{r}(\omega)=\left[
\begin{matrix}
r_{xx}(\omega) & r_{xy}(\omega) \\
-r_{xy}(\omega) & r_{xx}(\omega)
\end{matrix}\right]
\end{equation}
 \noindent Its eigenvalues are given accordingly by:
\begin{equation}\label{rpm}
r_{\pm}(\omega)= r_{xx}(\omega)\pm i r_{xy}(\omega) = \sqrt{R_{\pm}(\omega)}e^{i\theta_{\pm}(\omega)}.
\end{equation}
\noindent In the local limit they are related to the momentum-independent dielectric function via the Fresnel equations:
\begin{equation}\label{rpmFresnel}
r_{\pm}(\omega)=\frac{1-\sqrt{\epsilon_{\pm}(\omega)}}{1+\sqrt{\epsilon_{\pm}(\omega)}}.
\end{equation}
The magneto-reflectivity $R(\omega)$, unlike the zero-field reflectivity, depends on the polarization state of the incident light even if the sample is optically isotropic. For a circular polarization,  it takes one of the values $R_{\pm}(\omega)$ depending on the handedness. For the linear polarization, which can be represented as the sum of the two oppositely handed circular waves with equal amplitudes, it is given by the average value:
\begin{equation}
R(\omega) = \frac{R_+(\omega)+R_-(\omega)}{2}.
\label{R}
\end{equation}
\noindent For a general (elliptical) polarization, the reflectivity is an unequally weighted superposition of $R_{-}(\omega)$ and $R_{+}(\omega)$. Notably, many spectrometers, especially of the Fourier-transform type, generate light with an ill-defined and frequency-dependent polarization. This imposes placing an extra polarizing element in a magneto-optical experiment in order to have a well determined polarization. Below we shall consider the case where the incident light is linearly polarized and use Equation (\ref{R}).

It is obvious that the standard KK method  based on Equation (\ref{kkr}) does not apply since the reflectivity phases $\theta_{+}(\omega)$ and $\theta_{-}(\omega)$ are different. Moreover, it does not use the essential information coming from the Kerr angle:
\begin{equation}
\theta_{K}(\omega) = \frac{1}{2}\arg\left(\frac{r_-(\omega)}{r_+(\omega)}\right) = \frac{\theta_-(\omega)-\theta_+(\omega)}{2}.
\label{Theta}
\end{equation}

Here we prove mathematically that the complex reflectivities $r_{\pm}(\omega)$ can be model-independently extracted from $R(\omega)$ and $\theta_{K}(\omega)$ if these function are known on
the entire real-frequency axis. We first introduce an auxiliary complex function:
\begin{equation}
f(\omega) = \ln\left(\frac{r_-(\omega)}{r_+(\omega)}\right) = \frac{1}{2}\ln c(\omega) + 2 i\theta_K(\omega),
\label{functionf}
\end{equation}
\noindent where
\begin{equation}
c(\omega) = \frac{R_-(\omega)}{R_+(\omega)}
\label{dichroism}
\end{equation}
\noindent is the circular-dichroism ratio. As the imaginary part of $f(\omega) $ is the Kerr angle known from experiment, the circular dichroism can be computed via the KK transformation\cite{SmithJOSA76}:
\begin{equation}
c(\omega) = \exp\left\{\frac{8\omega}{\pi}\wp\int_0^\infty\frac{\theta_K(x)}{x^2-\omega^2}dx\right\}.
\label{kk1}
\end{equation}

\noindent Combining this result with Equation (\ref{R}) we can extract separately the reflectivity for each handedness:
\begin{equation}
R_{+}(\omega) = \frac{2R(\omega)}{1 + c(\omega)}, R_{-}(\omega) = \frac{2c(\omega)R(\omega)}{1 + c(\omega)}.
\label{RpRm}
\end{equation}

\noindent The sought phases can now be obtained by applying the KK transformation to $R_{+}(\omega)$ and $R_{-}(\omega)$:
 \begin{equation}\label{kk2}
\theta_{\pm}(\omega) = -\frac{1}{\pi}\wp\int_{-\infty}^{\infty}\frac{\ln
\sqrt{R_{\pm}(x)}}{x-\omega}dx + \pi,
\end{equation}
\noindent which can be reduced, using the parity relation $R_{\pm}(-\omega)=R_{\mp}(\omega)$, to:
\begin{equation}\label{thetamo}
\theta_{\pm}(\omega) = \tilde{\theta}(\omega)\mp \theta_{K}(\omega),
\end{equation}
\noindent where
\begin{equation}\label{thetastar}
\tilde{\theta}(\omega) =\frac{\theta_{+}(\omega)+\theta_{-}(\omega)}{2}= -\frac{\omega}{\pi}\wp\int_{0}^{\infty}\frac{\ln
\tilde{R}(x)}{x^2-\omega^2}dx + \pi
\end{equation}
\noindent and
\begin{equation}\label{Rstar}
\tilde{R}(\omega) = \sqrt{R_{+}(\omega)R_{-}(\omega)}=\frac{2\sqrt{c(\omega)}}{1+c(\omega)}R(\omega).
\end{equation}

Equations (\ref{kk1})-(\ref{Rstar}) constitute a self contained recipe to extract the complex MO reflectivity for each handedness and Equation (\ref{rpmFresnel}) allows computing the complex dielectric functions $\epsilon_{\pm}(\omega)$. An important implication of this result is that spectra of $R(\omega)$ and $\theta_{K}(\omega)$ measured in a broad spectral range and supplemented with proper extrapolations, bear \textit{sufficient information} for complete MO spectroscopy.

It is worth noting that the described algorithm relies on the non-trivial assumption that the functions $\ln r_{\pm}(\omega)$ are analytical, \textit{i.e.} they do not exhibit  poles or other singularities in the upper complex semiplane $\Im \omega > 0$. While this holds for normal reflectance from a bulk sample, the analyticity can break down, for example, for a partially transparent sample. In this case, the variational KK constrained analysis \cite{KuzmenkoRSI05} should instead be applied. In the next section we describe the application of the variation approach to a much broader set of magneto-optical experiments.

\section{Variational Kramers-Kronig analysis of magneto-optical spectra}
\label{SectionVariational}

First we remind the basic idea of the KK constrained variational analysis in zero field \cite{KuzmenkoRSI05}. The measured spectra such as reflectivity, transmission or any combination of them are initially fitted with a KK consistent and sum-rule satisfying model dielectric function. Most typically, this is a sum of a limited number of Drude-Lorentz oscillators:

\begin{equation}\label{DL}
\epsilon_{\mbox{\scriptsize mod}}(\omega) = \epsilon_{\infty} +
\sum_{k}\frac{\omega_{p,k}^2}{\omega_{0,k}^2 - \omega^2 -
i\gamma_{k}\omega },
\end{equation}
\noindent where each Lorentzian is parametrized by the oscillator frequency $\omega_{0,k}$ (equal to zero for a Drude component), the linewidth $\gamma_{k}$ and the plasma frequency $\omega_{p,k}$. The parameter $\epsilon_{\infty}$ is the contribution from the high-energy optical transitions. Once a reasonable fit is achieved, these parameters are fixed, and a so-called variational dielectric function (VDF) is added to the model:

\begin{equation}\label{VarPlusMod}
\epsilon(\omega) = \epsilon_{\mbox{\scriptsize mod}}(\omega) +  \epsilon_{\mbox{\scriptsize var}}(\omega).
\end{equation}

\noindent The VDF is defined on a dense mesh of the experimental frequency points $\omega_{i} $ $(i = 1, .., N) $, which usually coincide with the experimental datapoints or two times less dense (\emph{i.e. }every second datapoint). It is constructed as a sum of elementary KK-consistent functions $\epsilon_{i}^{\triangle}(\omega)$:

\begin{equation}\label{EpsVar}
    \epsilon_{\mbox{\scriptsize var}}(\omega)=\sum_{i=2}^{N-1}A_{i}\left[\epsilon_{i}^{\triangle}(\omega)+\epsilon_{i}^{\triangle*}(-\omega)\right],
\end{equation}
\noindent where the coefficients $A_{i} = \epsilon_{\mbox{\scriptsize var}}(\omega_{i})$ are regarded as free parameters. The imaginary part of $\epsilon_{i}^{\triangle}(\omega)$ has a triangular shape:
\begin{equation}\label{TriagEps2}
    \Im \epsilon_{i}^{\triangle}(\omega)=\left\{\begin{array}{cl}
      \frac{\omega-\omega_{i-1}}{\omega_{i}-\omega_{i-1}}& \mbox{, } \omega_{i-1}<\omega\leq\omega_{i} \\
      \frac{\omega_{i+1}-\omega}{\omega_{i+1}-\omega_{i}} & \mbox{, } \omega_{i}<\omega<\omega_{i+1} \\
      0                                               & \mbox{, } \mbox{otherwise}
    \end{array}\right.
\end{equation}
\noindent while its real part is obtained by the KK transformation:
\begin{eqnarray}\label{TriagEps1}
&&\Re \epsilon_{i}^{\triangle}(\omega)=\frac{1}{\pi}\wp\int_{-\infty}^{\infty}\frac{\Im \epsilon_{i}^{\triangle}(x)}{x-\omega}dx = \frac{1}{\pi}\left[ \frac{g(\omega-\omega_{i-1})}{\omega_{i}-\omega_{i-1}}\right.\nonumber\\
&-&\left.\frac{(\omega_{i+1}-\omega_{i-1})g(\omega-\omega_{i})}{(\omega_{i}-\omega_{i-1})(\omega_{i+1}-\omega_{i})}+\frac{g(\omega-\omega_{i+1})}{\omega_{i+1}-\omega_{i}}\right],
\end{eqnarray}
\noindent where $g(x)=x\ln|x|$. The second term to Equation (\ref{EpsVar}) serves to satisfy the physical condition $\epsilon(-\omega)=\epsilon^{*}(\omega)$. The entire collection of data is finally refitted by adjusting all $A_{i}$ values independently. As a rule, the fit quality is nearly perfect since the number of parameters is equal or comparable to the number of datapoints.
Since the variational method does not imply the analyticity of any other response function than $\epsilon(\omega)$ it applies to virtually any type of optical experiments.

It appears that this approach can also be applied to magneto-optics. The Drude-Lorentz model is modified in the magnetic field as follows\cite{}:
\begin{equation}
\epsilon_{\mbox{\scriptsize  mod},\pm}(\omega) = \epsilon_{\infty\pm} + \sum_{k}\frac{\omega_{p,k}^2}{\omega_{0,k}^2-\omega^2-i\gamma_{k}\omega\mp\omega_{c,k}\omega},
\label{magnetoDL}
\end{equation}

\noindent where a cyclotron frequency $\omega_{c,k}$ is introduced for each Lorentz term. At first one has to fit the entire set of MO data as close as possible using Equation (\ref{magnetoDL}) or any other KK-consistent and sum-rule compliant function. Then the model parameters should be fixed and a magneto-VDF should be added:
\begin{equation}\label{DL}
\epsilon_{\pm}(\omega) = \epsilon_{\mbox{\scriptsize mod}\pm}(\omega) +  \epsilon_{\mbox{\scriptsize var}\pm}(\omega)
\end{equation}

\noindent with the following structure:
\begin{equation}\label{VarPlusModMO}
\epsilon_{\mbox{\scriptsize var}\pm}(\omega)=\sum_{i=2}^{N-1}\left[A_{i\pm}\epsilon_{i}^{\triangle}(\omega)+A_{i\mp}\epsilon_{i}^{\triangle*}(-\omega)\right]
\end{equation}

\noindent that  satisfies the physical relation $\epsilon_{\pm}(-\omega)=\epsilon_{\mp}^{*}(\omega)$. One can see that now there are two adjustable parameters for every mesh frequency: $A_{i+}=\epsilon_{\mbox{\scriptsize var}+}(\omega_{i})$ and $A_{i-}=\epsilon_{\mbox{\scriptsize var}-}(\omega_{i})$. By refitting the data we obtain model-independently the complex valued $\epsilon_{+}(\omega)$ and $\epsilon_{-}(\omega)$ and thus achieve the goal of MOKKA.

\section{Experimental procedure}
\label{SectionExp}

In this section we shall describe an experiment to measure broadband magneto-reflectivity (magneto-transmission) and the MO rotation spectra that can be treated by MOKKA.
Although the measurement protocols described below were optimized for Fourier transform infrared (FTIR) spectroscopy, they can also be applied, with some modifications, to other types of broadband spectrometers. We assume that the experimental setup allows the inversion of the magnetic field direction. Hereafter until the end of the Section we shall refer to the reflectance and the Kerr rotation, although exactly the same experimental procedure applies to the transmission and the Faraday rotation.

As it was emphasized in Section \ref{SectionTheory}, in magneto-optics the polarization state of the incident light should be well defined even if the sample is isotropic in zero field. Since the spectrometer output may be uncontrollably polarized, a polarizer should be put before the sample. Additionally, the optical path between the sample and the detector may also introduce some depolarization effects, in which case placing a second polarizer (often called analyzer) after the sample is also required. Therefore we shall consider a setup, where both polarizers are present, and which allows measuring the reflectivity and the Kerr rotation within the same experiment. This configuration is similar to standard two-polarizer ellipsometry\cite{AzzamBashara}, except that here we consider normal incidence, while ellipsometry is typically done at large illumination angles. We shall assume that the polarizer is fixed and the analyzer is rotated, although the same result is valid for the case where the analyzer is fixed and the polarizer turns. Since the sample is isotropic, we can set the polarizer angle to zero without loss of generality. As it is shown in Appendix A, the intensity measured by the detector as a function of the analyzer angle $A$ is given by:
\begin{eqnarray}\label{IvsA}
&&I_{A}(\omega, B) = C_{A}(\omega,B)\nonumber\\
&\times&\frac{R(\omega,B)+\tilde{R}(\omega,B)\cos [2A -2\theta_K(\omega,B)]}{2},
\end{eqnarray}
\noindent where the functions $R(\omega,B)$, $\tilde{R}(\omega,B)$ and $\theta_{K}(\omega,B)$ are defined in Section \ref{SectionTheory} and the prefactor $C_{A}(\omega,B)$ is determined by all elements in the optical path from the source to the detector apart from the sample. It may depend on the analyzer angle due to depolarization effects in the detection channel and on the field because of the changes in detector sensitivity and all mechanical movements that it may induce.

If we replace the sample by a reference with $R = \tilde{R} = 1$ and $\theta_K = 0$ then the detector intensity will be given by the Malus law:
\begin{equation}\label{IrefvsA}
I_{A, ref}(\omega,B) = C_{A}(\omega,B)\cos^2A.
\end{equation}
\noindent Thus the reference measurement allows us to determine $C_{A}(\omega,B)$. \cite{Note2}. Since the choice of the best reference procedure is setup- and sample-dependent, we do not discuss this issue here\cite{Note3}.

Below we discuss two measurement protocols, both based on Equation (\ref{IvsA}) but adapted for different experimental situations.

\subsection{Complete protocol}

As it is common in polarimetric experiments, the most rigorous approach involves measuring the raw spectra $I_{A}(\omega,B)$ for a large number of analyzer angles spread equidistantly between $0^{\circ}$ and $360^{\circ}$.  The same should be done for the reference, in order to obtain the prefactor $C_{A}(\omega,B)$ as discussed above. For each frequency and each field value one can fit the function $F(A)= I_{A}/C_{A}$ by the formula $\alpha+\beta\cos (2A - 2A_m)$ with the fitting parameters $\alpha$, $\beta$ and $A_m$.  The reflectivity can be determined by the relation $R(\omega,B) = 2\alpha(\omega,B)$. The parameter $A_m$ formally corresponds to the Kerr angle. However, using the relation $\theta_K(\omega,B) =A_m(\omega,B)$ may result in a considerable systematic error due to polarizer imperfections. This problem is easily removed by subtracting the zero-field value: $\theta_K(\omega,B) = A_m(\omega,B)-A_m(\omega,0)$.

Our experience shows that measuring at the two polarities of the magnetic field improves both the systematic accuracy and the signal-to-noise ratio. Based on the parity relations $R(\omega,-B) = R(\omega,B) $ and $\theta_K(\omega,-B) = -\theta_K(\omega,B)$ the following equations can be derived:
\begin{eqnarray}\label{ReflThetaExpComplete}
R(\omega,B) &=& \alpha(\omega,B)+\alpha(\omega,-B),\\
\theta_K(\omega,B) &=& \frac{A_m(\omega,B)-A_m(\omega,-B)}{2}.
\end{eqnarray}
\noindent that we actually recommend to use.

The value of $\beta/\alpha = \tilde{R}(\omega)/R(\omega) =2\sqrt{c(\omega)}/(1+c(\omega))$ can in principle be utilized to determine the circular dichroism ratio $c(\omega)$. However, this only works well if the dicroism, and therefore the ellipticity of the reflected light, is strong enough, which is often not the case\cite{Note4}. Fortunately, the implementation of MOKKA does not require $c(\omega)$ since it is computed via a KK transformation of $\theta_K(\omega)$.

\subsection{Fast protocol}

The measurement scheme described above is rigorous but quite slow as it involves measuring the raw intensity spectra at dozens of analyzer angles for the sample and the reference. Fortunately, in many situations one can use an alternative protocol, which requires much fewer spectra to be measured. This method is accurate if the rotation angle is small (for example $\theta_K < 0.2$ rad), so that $\sin \theta_K\approx \theta_K$ and $\cos \theta_K\approx 1$. Since for small rotations the dichroism is also weak, $|c-1|\ll 1$, the ratio $\tilde{R}(\omega)/R(\omega)  =1 + O((c-1)^2)$ and hence we can ignore the small difference between $R(\omega)$ and $\tilde{R}(\omega)$.

The fast method hinges on the fact that the intensity is most sensitive to the MO rotation when the polarizers are oriented at $+45^{\circ}$ or $-45^{\circ}$ with respect to each other and is almost insensitive to it for $A = 0$, \emph{i.e.} when they are parallel. Within the approximations made above, Equation (\ref{IvsA}) gives the following detector intensity for these analyzer positions:
\begin{eqnarray}
I_{\pm45}(\omega,B) &\approx&C_{\pm45}(\omega,B)R(\omega,B)\frac{1\pm2\theta_K(\omega,B)}{2}\label{I45},\\
I_{0}(\omega,B)&\approx&C_{0}(\omega,B) R(\omega,B).\label{I0}
\end{eqnarray}
\noindent and for the reference:
\begin{eqnarray}
I_{\pm45,ref}(\omega,B) &\approx&\frac{C_{\pm45}(\omega,B)}{2}\label{I45ref},\\
I_{0,ref}(\omega,B)&\approx&C_{0}(\omega,B)\label{I0ref}.
\end{eqnarray}

An obvious way to determine the reflectivity is to use the measurement in the parallel polarizer configuration:
\begin{eqnarray}\label{RFast}
R(\omega, B) \approx \frac{I_{0}(\omega,B)}{I_{0,ref}(\omega,B)},
\end{eqnarray}
\noindent which has the additional advantage that the optical signal is at maximum.

In order to obtain the magneto-optical rotation, we first we assume that $C_{A}(\omega,+B)=C_{A}(\omega,-B)$. Then we have:
\begin{eqnarray}\label{Rho}
\rho_{\pm45}(\omega,B)=\frac{I_{\pm45}(\omega,+B)}{I_{\pm45}(\omega,-B)}\approx 1\pm4\theta_K(\omega,B),
\end{eqnarray}
\noindent where again we used that $\theta_K(\omega,-B) = -\theta_K(\omega,B)$. This suggests that the rotation can be obtained based on the measurement for only one of the angles $+45^{\circ}$ or $-45^{\circ}$, without reference:
\begin{eqnarray}\label{ThetaFast1}
\theta_{K}(\omega,B) \approx \pm\frac{\rho_{\pm45}(\omega,B)-1}{4}.
\end{eqnarray}
\noindent In this way the rotation can be measured even in experimental setups, where \emph{in situ} rotation of the analyzer is not possible as one can simply place the sample between two fixed polarizers mounted at $45^{\circ}$ with respect to each other.

If $C_{A}(\omega,+B)$ is not equal to $C_{A}(\omega,-B)$ then the procedure should be modified. The most obvious modification is to take the same measurement with a reference:
\begin{eqnarray}\label{RhoRef}
&&\rho_{\pm45,ref}(\omega,B)=\frac{I_{\pm45,ref}(\omega,+B)}{I_{\pm45,ref}(\omega,-B)}
\end{eqnarray}
\noindent and to use the corrected formula:
\begin{eqnarray}\label{ThetaFast2}
\theta_{K}(\omega,B) \approx \pm\frac{1}{4}\left[\frac{\rho_{\pm45}(\omega,B)}{\rho_{\pm45,ref}(\omega,B)}-1\right].
\end{eqnarray}

If the analyzer can be rotated then a more accurate result can be obtained by combining measurements at $+45^{\circ}$ and $-45^{\circ}$. One can use the formula:
\begin{eqnarray}\label{ThetaFast3}
\theta_{K}(\omega,B) \approx \frac{\rho_{+45}(\omega,B)-\rho_{-45}(\omega,B)}{8}
\end{eqnarray}
\noindent if $C_{A}(\omega,+B)=C_{A}(\omega,-B)$ or
\begin{eqnarray}\label{ThetaFast4}
\theta_{K}(\omega,B) \approx \frac{1}{8}\left[\frac{\rho_{+45}(\omega,B)}{\rho_{+45,ref}(\omega,B)}-\frac{\rho_{-45}(\omega,B)}{\rho_{-45,ref}(\omega,B)}\right]
\end{eqnarray}
\noindent otherwise. Equations (\ref{ThetaFast3}) and (\ref{ThetaFast4}) are less sensitive to polarizer imperfections than Equations (\ref{ThetaFast1}) and (\ref{ThetaFast2}). Additionally, one can show that Equation (\ref{ThetaFast3}) provides a more accurate result than Equation (\ref{ThetaFast1}) when a weak field asymmetry of $C_{A}(\omega,B)$ is present. This allows a reference-free measurement of the MO rotation without imposing stringent conditions on the stability of the optical elements with respect to the field.

\section{Application examples}
\label{SectionApp}

\subsection{Bismuth}

Bismuth is a canonical semimetal with a low carrier density ($\sim$10$^{17}$cm$^{-3}$), a long mean-free path and a small cyclotron mass. These properties, combined with a strong spin-orbit interaction, give rise to spectacular magneto-optical effects~\cite{BoylePR58,LaxPRL60,BurgielPR65,VerdunPRB76,VecchiPRB76}. While some classical phenomena, such as quantum oscillations \cite{ShubnikovdeHaas30,deHaasvanAlphen30}, were first discovered in bismuth, a number of new phenomena were recently found in it and in related compounds, such as a transition to a 3D topological insulating state upon Sb substitution~\cite{HsiehNature08}, a valley-ferromagnetic state~\cite{LiScience08}, a valley-nematic Fermi liquid state~\cite{ZhuNP11} and avoided Lifshitz-type semimetal-semiconductor transition~\cite{ArmitagePRL10}. Therefore this material makes an excellent case for applying MOKKA.

Bismuth has a rhombohedral (or trigonal) crystal lattice and therefore it is optically anisotropic. However, it is optically isotropic within the plane, which is perpendicular to the trigonal axis and therefore the described technique applies when the propagation of light and the magnetic field are along the trigonal axis. Fortunately, the material cleaves easily (in liquid nitrogen) in the same plane, making it possible to prepare a high-quality atomically flat surface of the needed orientation.

% crop:
% l: 1.4
% b: 1.7
% r:  0.95
% t: 2.85

\begin{figure}[htb]
\centering
\includegraphics[width=1\columnwidth]{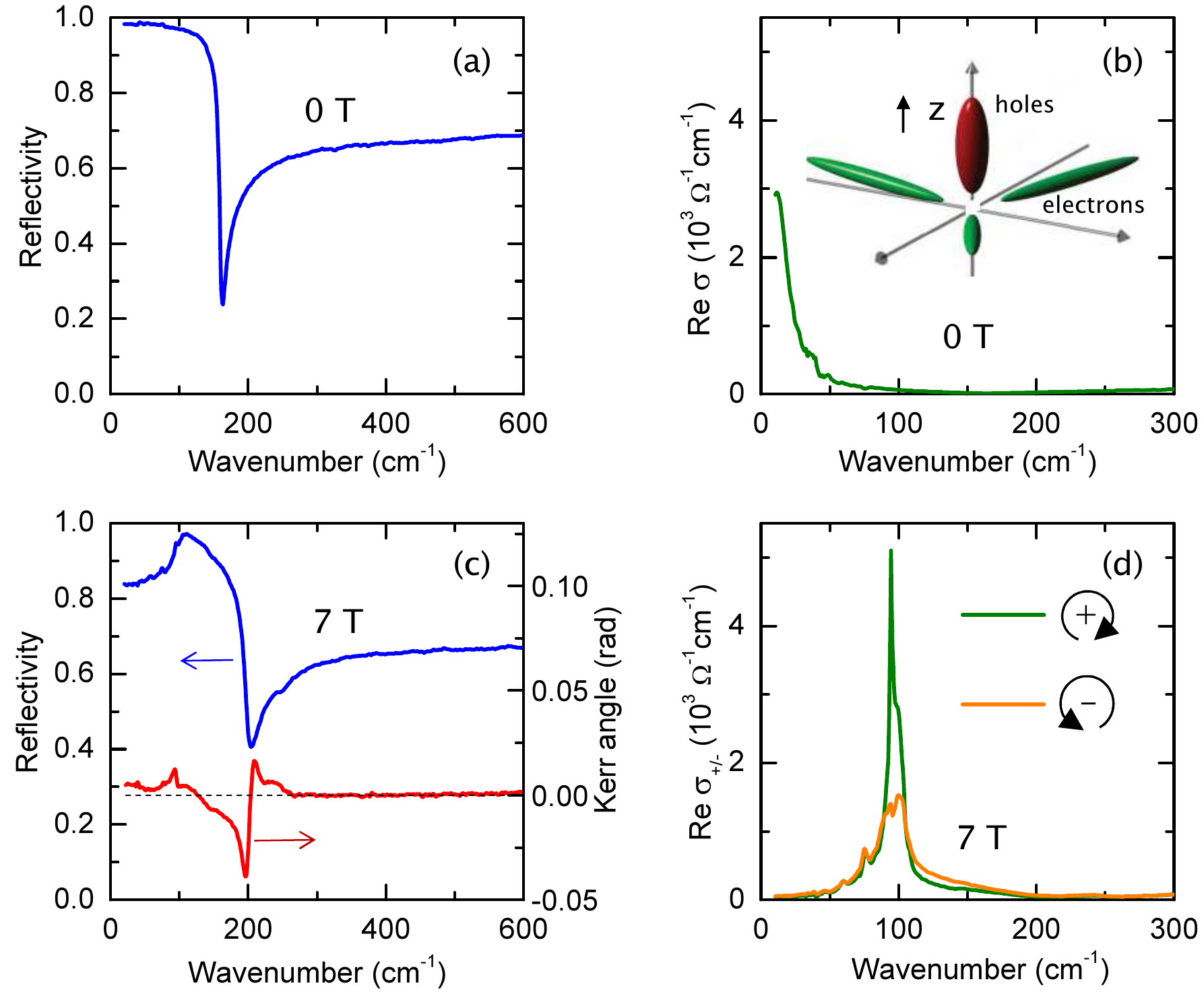}
\caption{Far-infrared optical and magneto-optical infrared spectra of bismuth at 10 K for the polarization perpendicular to the trigonal ($z$) axis. Panels (a) and (b) show the zero-field reflectivity and optical conductivity respectively. Panel (c) presents magneto-reflectivity (in blue) and Kerr rotation (in red) at 7 T. (d) The real (absorptive) part of the optical conductivity for the two circular polarizations obtained using MOKKA. The inset of panel (b) depicts the electron and hole Fermi pockets (adapted From Ref.\cite{ZhuPRB11}).}
\label{FigBismuth}
\end{figure}

Infrared magneto-reflectivity and Kerr angle spectra at nearly normal incidence ($\simeq 8^\circ$) were measured between 20 and 650 cm$^{-1}$ with the resolution of 1 cm$^{-1}$ at 10 K in a split-coil superconducting magnet attached to an FTIR spectrometer. Both protocols described in Section \ref{SectionExp} were used for comparison and they gave almost identical results. A Hg light source and a liquid He cooled bolometric detector were used. A crystal of 99.9999$\%$ pure Bi had a surface of $3\times5$ mm$^2$ and a thickness of 1 mm. Free standing wire-grid gold polarizers were used. The magnetic field was applied along the trigonal axis and therefore almost parallel to the propagation of light. The noise in the reflectivity spectra was less that 0.5 \%. The accuracy of the Kerr angle was better than 1 mrad.

The zero-field reflectivity, shown in Figure \ref{FigBismuth}a, features a sharp plasma edge at around 150 cm$^{-1}$ due to highly mobile charge carriers and it is in good agreement with previous works \cite{TediosiPRL07,LaForgePRB10}. The optical conductivity obtained by the usual KK analysis, is presented in Figure \ref{FigBismuth}b. It has a low-energy Drude peak and a linearly growing contribution from low-lying interband transitions.

The magneto-reflectivity and the Kerr angle spectra at 7 T are shown in Figure \ref{FigBismuth}c. One can note that the plasma edge is significantly blue-shifted by the field, which is a magnetoplasma effect\cite{PalikFurdyna}. The reflectivity also shows a strong decrease below 100 cm$^{-1}$, which can be attributed to the enhancement of the penetration depth and therefore an enhanced optical absorption below the cyclotron frequency. The Kerr rotation has a strong spectral structure clearly associated with the plasma edge at 200 cm$^{-1}$ and a weaker feature at 100 cm$^{-1}$, close to the frequency below which the reflectivity starts to decrease. Remarkably, the Kerr angle is essentially zero above 250 cm$^{-1}$.

Figure \ref{FigBismuth}d presents the optical conductivity $\sigma_\pm(\omega)$ for the left- and right-handed circular polarized light obtained by the variational magneto-optical KK analysis, where a Drude-Lorentz model with only a few terms was used as a starting point. Applying the direct integration procedure described in Section \ref{SectionTheory} provided essentially the same conductivity curves. The most prominent spectral structure in both polarizations is a absorption band at 80-110 cm$^{-1}$, which consists of several peaks. The spectral shape of this absorption is quite different for the two light helicities. The Fermi surface in bismuth is highly anisotropic. It features one hole pocket and three electron pockets with a Dirac-like band dispersion as shown in the inset of Figure \ref{FigBismuth}b. The presence of multiple cyclotron energies is obviously related to this complex band structure.

The physical interpretation of this data is beyond the scope of this paper and will be presented elsewhere \cite{LevalloisUnpublished}. Here we only note that MOKKA allows us to obtain the true shape of the MO conductivity and to determine the absolute spectral weight of the cyclotron resonance peaks for each handedness separately in a model-independent fashion.

\subsection{Graphite}

\begin{figure}[htb]
\centering
\includegraphics[width=1\columnwidth]{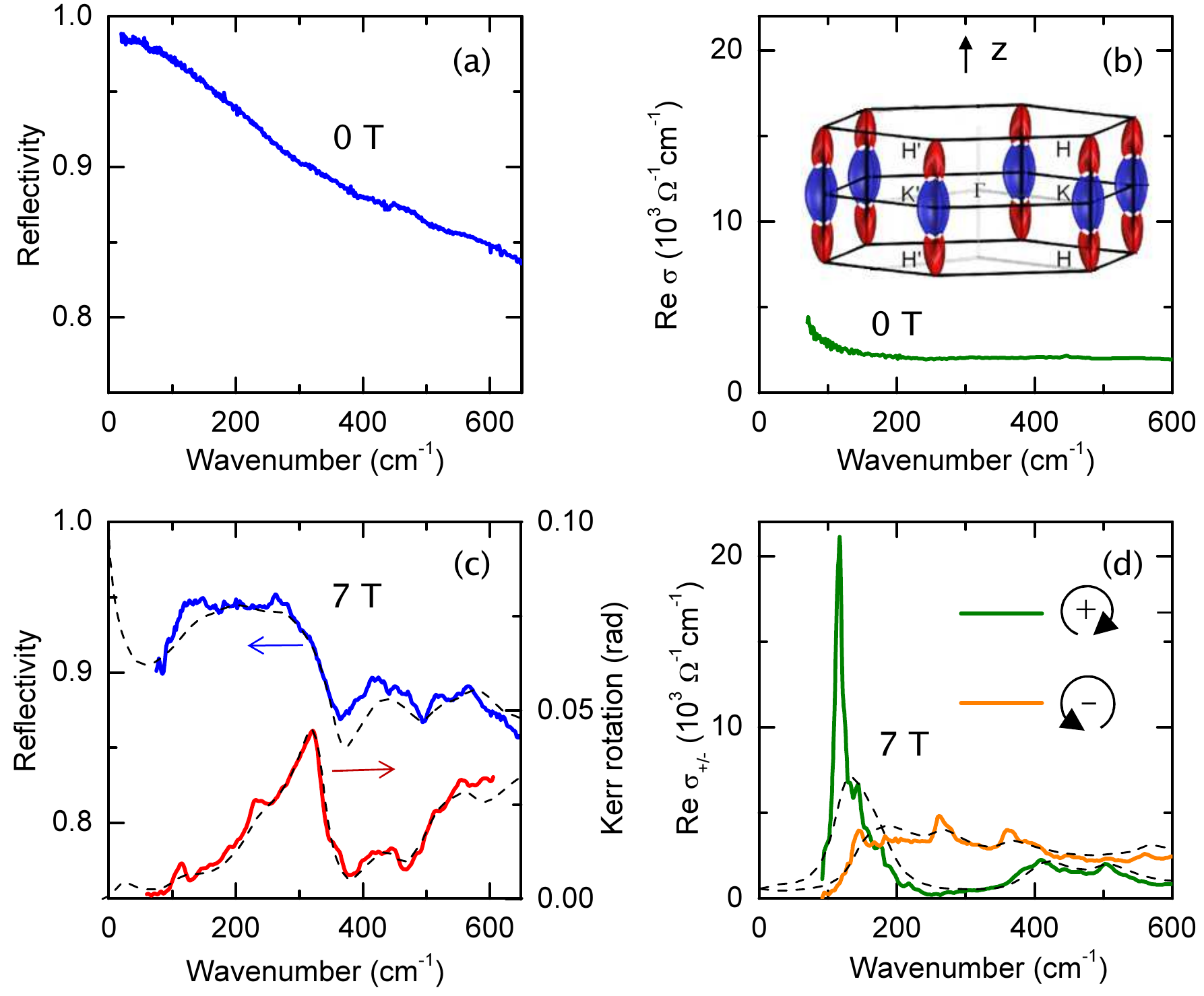}
\caption{Far-infrared optical and magneto-optical spectra of graphite at 10 K. Panels (a) and (b) show the zero-field reflectivity and optical conductivity respectively. Panel (c) presents magneto-reflectivity (in blue) and Kerr rotation (in red) at 7 T. (d) The real (absorptive) part of the optical conductivity for the two circular polarizations obtained using MOKKA. The dashed curves in Panels (c) and (d) correspond to the tight-binding Hamiltonian fits\cite{LevalloisSSC12}. The inset depicts the Fermi surface (electrons pockets in blue, hole pockets in red), adapted From Ref.\cite{SchneiderThesis}).}
\label{FigGraphite}
\end{figure}

Graphite is another textbook semimetal, where several physical phenomena were discovered, and which is additionally of a tremendous practical importance. In the last decade it again attracted much interest in the research community because of the exfoliation of graphene\cite{NovoselovScience04}. Graphite is composed of weakly coupled honeycomb carbon layers with predominantly Bernal stacking and is therefore extremely anisotropic. The tiny Fermi surface consists of trigonally warped electron and hole pockets stretched along the K-H corner of the hexagonal Brillouin zone, as shown in the inset of Figure \ref{FigGraphite}b. The very peculiar optical and magneto-optical properties of graphite were widely studied \cite{GaltPR56,ToyPRB77,TaftPhilipPR65,LiPRB06,KuzmenkoPRL08,LevalloisSSC12}. Recently, we measured infrared magneto-reflectivity and Kerr rotation spectra and fitted them using a tight-binding model Hamiltonian\cite{LevalloisSSC12}. Here we shall demonstrate a model-independent extraction of handedness-resolved magneto-optical conductivity from the same data.

% crop:
% l: 1.4
% b: 1.7
% r:  0.95
% t: 2.85

Figures \ref{FigGraphite}a and \ref{FigGraphite}b present the far-infrared region of the zero-field reflectivity and the KK-derived optical conductivity of highly-ordered pyrolytic graphite (HOPG) at 10 K\cite{KuzmenkoPRL08}. The reflectivity is growing monotonously towards unity at zero frequency and the optical conductivity is essentially constant, apart from the lowest frequencies, where a small Drude peak is observed. Above 200 cm$^{-1}$ the conductivity is close to $\sigma_{0} = e^2/4\hbar$ per layer ($\approx$ 1820 $\Omega^{-1}$cm$^{-1}$), which is the theoretically expected conductivity of monolayer graphene due to optical transitions between conical valence and conduction bands\cite{KuzmenkoPRL08}.

Figure \ref{FigGraphite}c shows the experimental results at 7 T. The reflectivity modifies dramatically in field showing intense spectral structures owing to the formation of Landau levels. The Kerr rotation also shows a strong and rather nontrivial spectral dependence.  We successfully modeled these data in fields from 0 to 7 T using a Slonczewski-Weiss-McClure Hamiltonian with Landau-level quantization and the Kubo formula as described in Ref. \onlinecite{LevalloisSSC12}. The fits at 7 T are shown as dashed lines. Although very good, they show some deviations from the experiment and therefore do not provide the precise shape of the optical conductivity. Here we use these physically meaningful fits, instead of the phenomenological Drude-Lorentz model, with fixed parameters as a starting point and apply a magneto-VDF as discussed in Section \ref{SectionVariational}. This allows us to achieve a nearly perfect fit (not shown) and calculate the optical conductivity for the two circular polarizations shown in Figure \ref{FigGraphite}d. One can see that the true optical conductivity shows much sharper peaks than the model. This is likely due to the fact that the Kubo formula used in Ref. \onlinecite{LevalloisSSC12} was assuming a frequency-independent scattering rate, thus strongly overestimating the optical width of low-energy LL transitions.

In contrast to the case of bismuth, the two conductivity curves of graphite are markedly different. This is explained by non-trivial magneto-optical selection rules in graphite that are dominated by the trigonal warping, as discussed in details in Ref.\onlinecite{LevalloisSSC12}.

\subsection{Multilayer graphene on SiC}

\begin{figure}[htb]
\centering
\includegraphics[width=1\columnwidth]{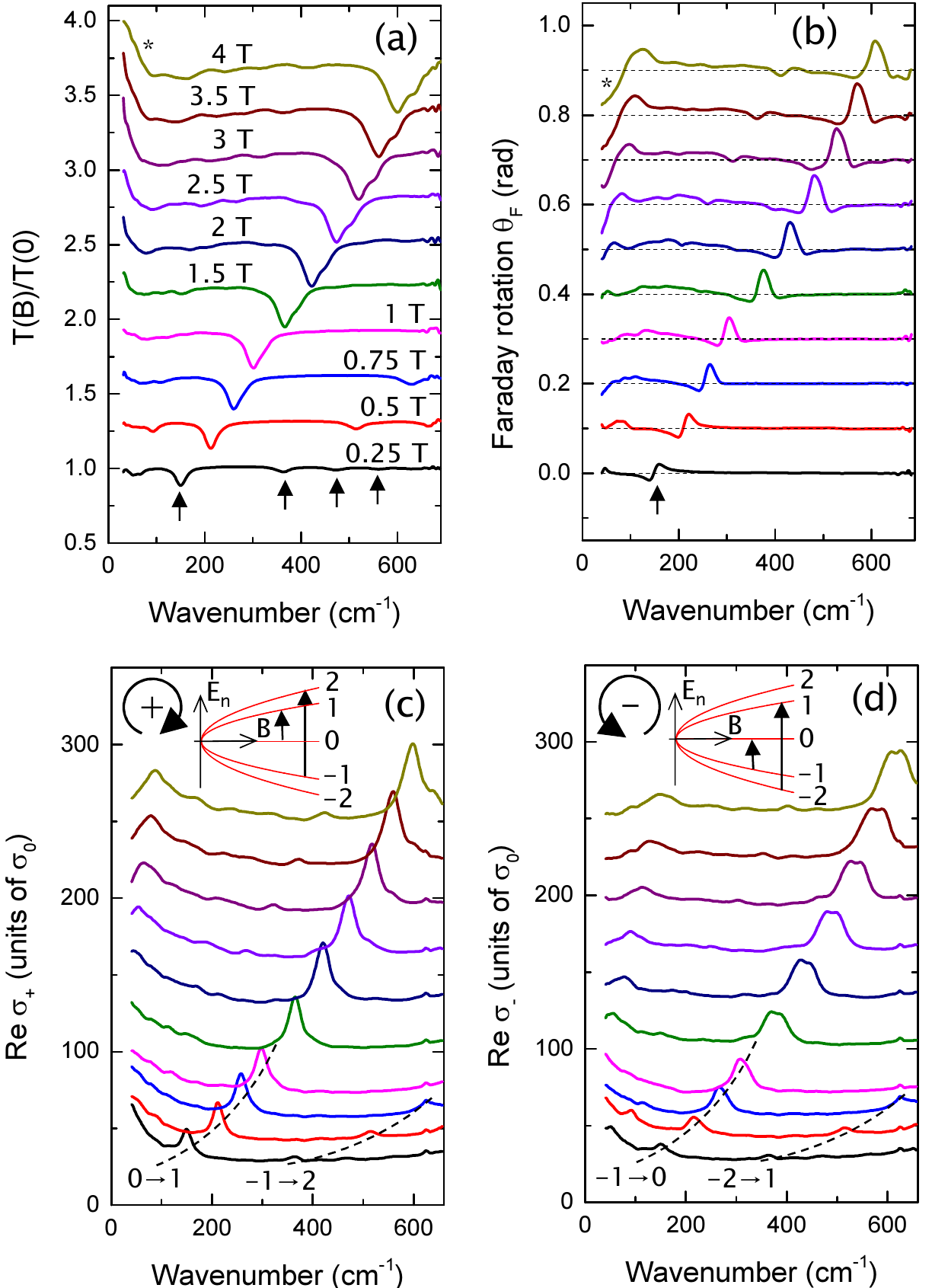}
\caption{Far infrared magneto-transmission (a) and Faraday rotation (b) of multilayer graphene on the Si-face of 6H-SiC at 5 K at several values of perpendicular magnetic field between 0.25 T and 4 T. Arrows indicate the LL transitions from neutral layers, the asterisk shows the Drude contribution from highly doped layers. Panels (c) and (d) show the two-dimensional optical conductivity normalized by $\sigma_{0} = e^2/4\hbar$ for circular light of each handedness obtained using variational MOKKA. The corresponding Landau level transitions are sketched in the insets. In all graphs, the curves are shifted vertically for clarity, except at 0.25 T. The zero levels for the Faraday rotation curves are shown by the dashed lines.}
\label{FigGraphene}
\end{figure}

As a third example, we show an extraction of handedness-resolved magneto-optical conductivity of an ultrathin film on a MO-inactive substrate by applying MOKKA to a combination of magneto-transmission and Faraday rotation. Figures \ref{FigGraphene}a and \ref{FigGraphene}b present far-infrared spectra of $T(B)/T(0)$ and $\theta_F$ of multilayer graphene grown on the C-face of 6H-SiC \cite{BergerJPCB04} for several fields between 0.25 and 4 T at 5 K. The Fabry-Perot oscillations in the substrate were intentionally suppressed by reducing the spectral resolution to 5 cm$^{-1}$.

The two-dimensional (sheet) optical conductivities $\sigma_{\pm}(\omega)$ of graphene (Figures \ref{FigGraphene}c and \ref{FigGraphene}d)) were calculated model-independently using the variational MOKKA introduced in section \ref{SectionVariational} and the Fresnel formulas for the case of incoherent internal substrate reflections as detailed in Appendix B.

% crop:
% l: 0.7
% b: 1.3
% r:  0.8
% t: 5.5

The transmission features several absorption dips (marked by arrows for $B$=0.25 T) corresponding to optical transitions between Landau levels, in agreement with previous measurements \cite{SadowskiPRL06,CrasseeNP11}. The transition energies are proportional to $\sqrt{B}$ and match almost perfectly the expected values in monolayer graphene, where the LL energies are given by the formula:
\begin{equation}
E_{n} = \mbox{sign}(n)\sqrt{2\hbar e v_F^2|nB|}.
\end{equation}
\noindent Here $v_F\approx 10^6$m/s is the Fermi velocity and $n = 0, \pm 1 , \pm 2,  .. $ is the LL quantum number. It was argued \cite{SadowskiPRL06} that the monolayer-like dependence is due to a twisted stacking of the layers which decouples them from each other\cite{dosSantosPRL07} much more efficiently than in the Bernal-stacked graphite. The absolute spectral weight of the LL absorption is in fact much smaller than the expected value according to the number of layer (about 20 in this sample) for reasons that are not fully understood at the moment. In addition to quantum Landau level transitions, there is a structure at low energies (shown by the asterisk for 4 T) stemming from the cyclotron resonance of free carriers in strongly doped layers close to the substrate.

While the LL transitions $-n\rightarrow n+1$ and $-n-1\rightarrow n$ have the same energy, they are excited by circular polarizations of light \cite{OrlitaPotemskiReview} as sketched in the insets of Figures \ref{FigGraphene}c and \ref{FigGraphene}d). In a simple magneto-transmission experiment they unavoidably overlap. The advantage of MOKKA over is that it allows an unambiguous separation of transition with different handedness. In the presented case, one can clearly see that the shape of the peaks with the corresponding energy is quite different in  $\sigma_{+}(\omega)$ and $\sigma_{-}(\omega)$. For example, the peaks in $\sigma_{+}(\omega)$ are sharper. One can also notice that at low fields (0.25 and 0.5 T) the intensity of the $0\rightarrow 1$ peak is significantly larger than that of the $-1\rightarrow 0$ peak. A discussion of this electron-hole asymmetry was given in Ref. \cite{CrasseePRB11}.

\section{Summary}
\label{SectionSummary}

The goal of this paper is to introduce a technically simple solution for a complete-tensor broadband magneto-optical spectroscopy. The pathway that we propose is to perform an ellipsometry-like measurement at normal incidence with two polarizers and to supplement it with a properly generalized Kramers-Kronig analysis. We have proven mathematically that in the case of a bulk sample a combination of broadband magneto-reflectivity and Kerr angle that can be measured using this simple technique, allow a unique and model-independent extraction of handedness resolved dielectric function. Furthermore, we extended the variational Kramers-Kronig analysis\cite{KuzmenkoRSI05} to magneto-optical spectra, which can be applied in a variety of experimental situations, including transmission and Faraday rotation measurements on thin films deposited on a substrate as well as reflectivity on semi-transparent samples. This method can be also regarded as an optical-Hall type technique since it provides access both to the longitudinal and the Hall components of the optical conductivity.

As the examples of bismuth, graphite and epitaxial graphene on SiC have clearly shown, handedness resolved MO spectroscopy is especially useful for the systems, where both electrons and holes are present. In contrast to the often used approach, where the measured spectra (including Kerr or Faraday angles) are only compared to model curves\cite{SalghettiSSC99,OppeneerReview,HancockPRL11,LevalloisSSC12}, our technique allows extracting the true shape of the MO dielectric function and optical conductivity. This is important, where such a shape is difficult to reproduce theoretically, for example, due to a frequency-dependent scattering rate. Although all spectra presented in this paper were measured in the far-infrared region, the same technique can be obviously applied to any other optical range.

Further extensions of this method to more complicated situations can be envisaged. First, in the case, where the magnetic and magneto-electric susceptibility play a role one would need to increase the number of independent measurements. For example, measuring magneto-transmission and Faraday rotation for several films of different thickness and properly modifying the Fresnel equations may allow an independent extraction of $\epsilon_{\pm}(\omega)$ and  $\mu_{\pm}(\omega)$. Second, if non-local effects are essential then one should introduce the momentum dependence to the dielectric function and replace the Fresnel equations with more general electromagnetic relations. Remarkably, even in this case the complex reflectivity for each handedness can be extracted using the formalism presented in Section \ref{SectionTheory}, which does not rely on the Fresnel equations. Third, if a sample is anisotropic in zero field and the MO tensors $\hat{\epsilon}(\omega)$ and $\hat{r}(\omega)$ no longer satisfy the symmetry of Equations (\ref{condtensor}) and (\ref{rtensor}) the measurement technique has to be modified to involve rotating of both polarizers. As a final remark, the time-domain polarimetric spectroscopy \cite{vanMechelenPRL11,HancockPRL11,NeshatOE12} offers further opportunities when combined with the method described in this paper, as it allows measuring more independent quantities simultaneously.

\subsection*{Acknowledgments}
We thank M. Tran, N. Ubrig, D. van der Marel, J.-M. Poumirol and A.Akrap (University of Geneva) for valuable discussions. We are grateful to C. Uher (University of Michigan) and T. Seyller (University of Chemnitz) for providing us high-quality samples of bismuth and epitaxial graphene on SiC. This work was supported by the Swiss National Science Foundation and EU Graphene Flagship (contract no. CNECT-ICT-604391).

\section*{Appendix A}

\begin{figure}[htb]
\centering
\includegraphics[width=1\columnwidth]{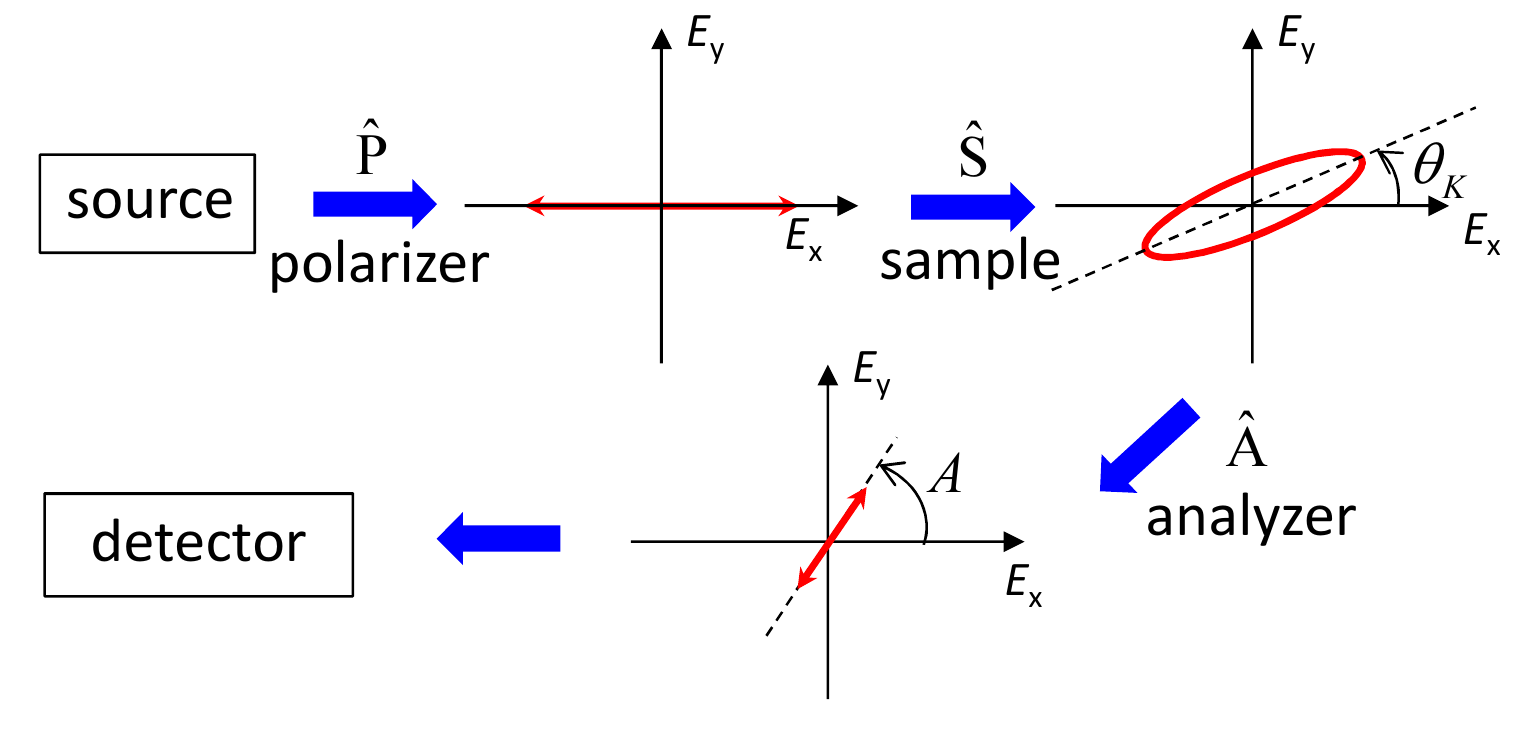}
\caption{Transformations of the polarization of light in the two-polarizer optical setup used for the magneto-optical experiment proposed in this paper.}
\label{FigJones}
\end{figure}

Here we use the Jones-matrix formalism to derive Equation (\ref{IvsA}). For simplicity, we shall refer to the reflection experiment, although exactly the same derivation applies to transmission. Below we consider the case where the first polarizer is fixed and the second polarizer (analyzer) is rotating. However, the same result applies when the first polarizer is rotated and the second one is fixed. The transformation of the polarization of light along the optical path is sketched in Figure \ref{FigJones}.

% crop:
% l: 2.85
% b: 1.0
% r:  2.5
% t: 4.6

Without loss of generality, we assume that the polarizer is aligned along the $x$ axis, so that its Jones matrix is
\begin{equation}\label{JonesP}
\hat{P}=t_{P}\begin{pmatrix}
          1 & 0 \\
          0 & 0 \\
        \end{pmatrix}
\end{equation}

\noindent where $t_{P}$ is the polarization transmission for the wanted polarization. This results in the following Jones vector of the radiation incident to the sample:

\begin{equation}\label{JonesEin}
\vec{E}_{in}=\hat{P}\vec{E}_{source}=\begin{pmatrix}  1 \\  0 \\ \end{pmatrix} t_{P} E_{0},
\end{equation}

\noindent where $E_{0}$ is the $x$ component of the amplitude of light leaving the spectrometer.
Using the Jones matrix of the sample
\begin{equation}\label{JonesS}
\hat{S}=\begin{pmatrix}
          r_{xx} & r_{xy} \\
          -r_{xy} & r_{xx} \\
        \end{pmatrix}
\end{equation}
\noindent we obtain the Jones vector of the reflected light:
\begin{equation}\label{JonesEref}
\vec{E}_{ref}=\hat{S}\vec{E}_{in}=\begin{pmatrix}  r_{xx} \\  -r_{xy} \\ \end{pmatrix} t_{P} E_{0},
\end{equation}
\noindent which in general corresponds to an elliptical polarization. The Jones matrix of the analyzer set at the angle  $A$ with respect to the $x$ axis is:
\begin{equation}\label{JonesS}
\hat{A}=t_{A} \begin{pmatrix}
          \cos^2 A & \cos A \sin A\\
          \cos A \sin A & \sin^2 A \\
        \end{pmatrix},
\end{equation}

\noindent where $t_{A}$ is the analyzer transmission. After the analyzer the following polarization state is achieved:
\begin{eqnarray}\label{JonesEout}
\vec{E}_{out}&=&\hat{A}\vec{E}_{ref}\nonumber\\
&=&\begin{pmatrix}  \cos A\\  \sin A \\ \end{pmatrix} \left(r_{xx} \cos A - r_{xy} \sin A\right)t_{A} t_{P} E_{0},
\end{eqnarray}

The signal at the detector is finally given by
\begin{equation}\label{Idet}
I_A(\omega) = \zeta\left|\vec{E}_{out}\right|^2 = C_{A}\left|r_{xx} \cos A - r_{xy} \sin A\right|^2
\end{equation}

\noindent where $\zeta$ is the cumulated transmission coefficient of all optical elements between the analyzer and the detector and $C_{A} = \zeta|t_{A} t_{P} E_{0}|^2$. Note that $\zeta$ and $C_{A}$ may depend on $A$ if some of these elements are polarization sensitive. In the following, we take into account that
\begin{eqnarray}\label{JonesRxxRxy}
r_{xx} = \frac{\sqrt{R_{+}}e^{-i\theta_{K}}+\sqrt{R_{-}}e^{i\theta_{K}}}{2}e^{i(\theta_{+}+\theta_{K})}\\
r_{xy} = \frac{\sqrt{R_{+}}e^{-i\theta_{K}}-\sqrt{R_{-}}e^{i\theta_{K}}}{2i}e^{i(\theta_{+}+\theta_{K})}
\end{eqnarray}
\noindent as it follows from Equation (\ref{rpm}). Substituting this into Equation (\ref{Idet}) yields the dependence of the detected signal on the analyzer angle:
\begin{eqnarray}\label{Idet2}
I_A = \frac{C_{A}}{4}\left| (\sqrt{R_{+}}e^{-i\theta_{K}}+\sqrt{R_{-}}e^{i\theta_{K}})\cos A\right.\nonumber\\
\left. + i(\sqrt{R_{+}}e^{-i\theta_{K}}-\sqrt{R_{-}}e^{i\theta_{K}})\sin A\right|^2
\end{eqnarray}

\noindent After simple algebra this reduces to:
\begin{eqnarray}\label{Idet3}
I_{A} = C_{A}\frac{R+\tilde{R}\cos (2A-2\theta_{K})}{2} .
\end{eqnarray}

This result has a straightforward geometrical interpretation. The detector signal is minimized when the analyzer is orthogonal to the main axis of the ellipse drawn by the electric-field vector of light reflected from the sample (Figure \ref{FigJones}). On the other hand, the orientation of this axis is given precisely by the Kerr angle.

\section*{Appendix B}

For the reference purpose, here we present equations needed to calculate the magneto-transmission $T = (T_{+} + T_{-})/2$ and the Faraday rotation $\theta_F$ of a MO-active film with the two-dimensional optical conductivity $\sigma_{\pm}$ deposited on a MO-inactive substrate with the complex refractive index $n_{s}$ and thickness $d_{s}$. The result depends strongly on whether or not the multiple internal reflections inside the substrate are taken into account. If they are negligible, which is, for example, the case for a strongly absorbing substrate, then the transmission normalized to the bare identical substrate and Faraday rotation are given by:

\begin{eqnarray}\label{TpmThetaFZero}
T_{\pm}^{0} &=& \left|1 + \frac{Z_{0}\sigma_{\pm}}{n_{s}+1}\right|^{-2},\\
\theta_F^{0} &=& \frac{1}{2}\arg\left(\frac{n_{s}+1+Z_{0}\sigma_{+}}{n_{s}+1+Z_{0}\sigma_{-}}\right),
\end{eqnarray}

\noindent where $Z_{0} $ is the vacuum impedance. In this situation the effect of the substrate as compared to a free standing film is to reduce the effective value of the film conductivity by $(n_s + 1)/2$ due to a screening of the electric field experienced by the film.

If multiple reflections contribute significantly to the signal then the effect of the substrate is more complex and is dependent on the degree of their mutual phase coherence. Decoherence can be caused by wedging of the substrate, a distribution of the angles of incidence, a finite resolution of the spectrometer and other factors. While it is in general quite hard to reproduce this effect quantitatively, the analytical result can be easily found in the two limiting cases. If the internal reflections are fully coherent then

\begin{eqnarray}\label{TpmThetaFCoh}
T_{\pm}^{c} &=& T_{\pm}^{0} \times\left|\frac{1 - p^2}{1 - pq_{\pm}}\right|^2, \\
\theta_F^{c} &=& \theta_F^{0}+ \frac{1}{2}\arg\left(\frac{1-pq_{+}}{1-pq_{-}}\right),
\end{eqnarray}

\noindent where
\begin{eqnarray}
p &=& \frac{n_{s}-1}{n_{s}+1}\exp{\left\{i\frac{\omega}{c} n_s d_s\right\}},\\
q_{\pm} &=& \frac{n_{s}-1-Z_{0}\sigma_{\pm}}{n_{s}+1+Z_{0}\sigma_{\pm}}\exp{\left\{i\frac{\omega}{c} n_s d_s\right\}}.
\end{eqnarray}

\noindent and the Fabry-Perot effect is present both in the transmission and the Faraday rotation \cite{UbrigOE13}. In the opposite case of the full incoherence \cite{CrasseeThesis} one obtains:

\begin{eqnarray}\label{TpmThetaFIncoh}
T_{\pm}^{i} &=& T_{\pm}^{0} \times\frac{1 - |p|^4}{1 - |pq_{\pm}|^2},\\
\theta_F^{i} &=&\theta_F^{0}+ \frac{1}{2}\arg\left(1-|p|^2 q_{-}^{*}q_{+}\right).
\end{eqnarray}

Interestingly, at selected frequencies the Fabry-Perot effect can increase the transmission and the Faraday rotation simultaneously as compared to the incoherent case, which can be important for MO applications \cite{UbrigOE13}. On the other hand, applying MOKKA is easier if the Fabry-Perot oscillations are fully suppressed.

%\bibliographystyle{apsrev4-1}
%\bibliography{biblio}

\section*{References}

\begin{thebibliography}{99}

\bibitem{FaradayPTRS1846} M. Faraday, Phil. Trans. R. Soc. \textbf{136}, 104 (1846).
\bibitem{KerrPM1877} J. Kerr, Phil. Mag. \textbf{3}, 321 (1877).
\bibitem{OnsagerPR31} L. Onsager, Phys. Rev. \textbf{37}, 405 (1931).
\bibitem{KramersNature26} H. A. Kramers, Nature {\bf 117}, 775 (1926).
\bibitem{KronigJOSA26} R. de Kronig, J. Opt. Soc. Am. {\bf 12}, 547 (1926).
\bibitem{LandauLifshitz} L. D. Landau and E. M. Lifshitz, {\it Electrodynamics of Continuous Media}, Oxford, Pergamon Press (1975).
\bibitem{DresselGruner} M. Dressel and G. Gr\"uner, {\it Electrodynamics of Solids}, Cambridge University Press (2002).
\bibitem{Note1} While the first equality is universal, the second one is valid only in the local limit, which is however widely used in material optics.
\bibitem{JahodaPR57} F. C. Jahoda, Phys. Rev. {\bf 107}, 1261 (1957).
\bibitem{MassonOL06} J.-B. Masson and G. Gallot, Optics Lett. {\bf 31}, 265 (2006).
%\bibitem{KaveevAO13} A. K. Kaveev, G. I. Kropotov, E. V. Tsygankova, I. A. Tzibizov, S. D. Ganichev, S. N. Danilov, P. Olbrich, C. Zoth, E. G. Kaveeva, A. I. Zhdanov, A. A. Ivanov, R. Z. Deyanov, and B. Redlich, Appl. Optics {\bf 52}, B60 (2013).
\bibitem{KaveevAO13} A. K. Kaveev \emph{et al}, Appl. Optics {\bf 52}, B60 (2013).
\bibitem{KarraiPRL92} K. Karrai, E. Choi, F. Dunmore, S. Liu, X. Ying, Qi Li, T. Venkatesan, H. D. Drew, Qi Li, and D. B. Fenner, Phys. Rev. Lett.  {\bf 69}, 355 (1992).
\bibitem{KangTSF11}  T. D. Kang, E. Standard, G. L. Carr, T. Zhou, M. Kotelyanskii, and A. A. Sirenko, Thin Solid Films {\bf 519}, 2698 (2011).
\bibitem{XiIPT14} X. Xi, R. J. Smith, T. N. Stanislavchuk, A. A. Sirenko, S. N. Gilbert, J. J. Tu, G. L. Carr, Infrared Phys. Tech. {\bf 67}, 436 (2014).
\bibitem{SatoJJAP81}  K. Sato, Jpn. J. Appl. Phys. {\bf 20}, 2403 (1981).
\bibitem{YamaguchiPRB98} S. Yamaguchi, Y. Okimoto, K. Ishibashi, and Y. Tokura, Phys. Rev. B \textbf{58}, 6862 (1998).
\bibitem{KuhneRSI14} P.K\"{u}hne, C. M. Herzinger, M. Schubert, J.A. Woollam, and T. Hofmann, Rev. Sci. Instrum. {\bf 85}, 071301 (2014).
\bibitem{SieversPR63} A.J. Sievers and M. Tinkham, Phys. Rev. \textbf{129}, 1566 (1963).
\bibitem{PimenovNP06} A. Pimenov, A. A. Mukhin, V. Yu. Ivanov, V. D. Travkin, A. M. Balbashov, and A. Loidl, Nature Physics \textbf{2}, 97 (2006).
\bibitem{QiPRB08} X.-L. Qi, T.L. Hughes and S.-C. Zhang, Phys. Rev. B  {\bf 78}, 195424 (2008).
\bibitem{KuzmenkoRSI05} A.B. Kuzmenko, Rev. Sci. Instrum. {\bf 76}, 083108 (2005).
\bibitem{AzzamBashara} R.M.A. Azzam and N. M. Bashara, {\it Ellipsometry and polarized light}, North-Holland (1977).
\bibitem{Note2} While $C_{A}(\omega,B)$ can be determined as $I_{A,ref}(\omega,B)/\cos^2A$, a better way to do obtain it is by fitting $I{ref}(A)$ while assuming a reasonable dependence of $C(A)$, for example, $C(A) = C_1 + C_2 \cos A$.
\bibitem{Note3} For a small or an irregularly shaped sample an evaporation of a thin metallic film on the surface may be required to obtain the correct absolute reflection.
\bibitem{Note4} This problem is analogous to the poor sensitivity of two-polarizer ellipsometry to a weak absorption resulting in a small ellipticity of reflected light. In ellipsometry, this can be solved by using retarders\cite{AzzamBashara}. Here we avoid this for the reasons mentioned in the introduction.
\bibitem{SmithJOSA76} D. Y. Smith, J. Opt. Soc. Am. {\bf 66}, 454 (1976).
\bibitem{BoylePR58} W. S. Boyle, A. D. Brailsford, and J. K. Galt, Phys. Rev. \textbf{109}, 1396 (1958).
\bibitem{LaxPRL60} B. Lax, J. G. Mavroides, H. J. Zeiger, and R. J. Keyes, Phys. Rev. Lett. \textbf{5}, 241 (1960).
\bibitem{BurgielPR65} J. C. Burgiel and L. C. Hebel, Phys. Rev. \textbf{140}, A925 (1965).
\bibitem{VerdunPRB76} H. R. Verd\'{u}n and H. D. Drew, Phys. Rev. B \textbf{14}, 1370 (1976).
\bibitem{VecchiPRB76} M. P. Vecchi, J. R. Pereira, and M. S. Dresselhaus, Phys. Rev. B \textbf{14}, 298 (1976).
\bibitem{ShubnikovdeHaas30} L. Shubnikov and W.Y. de Haas, Proc. Acad. Sci. Amsterdam, \textbf{33} 130 (1930).
\bibitem{deHaasvanAlphen30} W.Y. de Haas and P.M. van Alphen, Proc. Acad. Sci. Amsterdam \textbf{33} 1106 (1930).
\bibitem{HsiehNature08} D. Hsieh, D. Qian, L. Wray, Y. Xia, Y. S. Hor, R. J. Cava, and M. Z. Hasan, Nature \textbf{452}, 970 (2007).
\bibitem{ArmitagePRL10} N. P. Armitage, R. Tediosi, F. L\'{e}vy, E. Giannini, L. Forro and D. van der Marel, Phys. Rev. Lett. \textbf{104}, 237401 (2010).
\bibitem{LiScience08} L. Li, J. G. Checkelsky, Y. S. Hor, C. Uher, A. F. Hebard, R. J. Cava, and N. P. Ong, Science 321, \textbf{547} (2008).
\bibitem{ZhuNP11} Z. Zhu, A. Collaudin, B. Fauqu\'{e}, W. Kang, and K. Behnia, Nature Physics \textbf{8}, 89 (2011).
\bibitem{ZhuPRB11} Z. Zhu, B. Fauqu\'{e}, Y. Fuseya, and K. Behnia, Phys. Rev. B \textbf{84}, 115137 (2011).
\bibitem{TediosiPRL07} R. Tediosi, N. P. Armitage, E. Giannini, and D. van der Marel, Phys. Rev. Lett. \textbf{99}, 016406 (2007).
\bibitem{LaForgePRB10} A. D. LaForge, A. Frenzel, B. C. Pursley, T. Lin, X. Liu, J. Shi, and D. N. Basov, Phys. Rev. B \textbf{81}, 125120 (2010).
\bibitem{PalikFurdyna} E.D. Palik, J.K. Furdyna, Rep. Prog. Phys. \textbf{33}, 1193 (1970).
\bibitem{LevalloisUnpublished} J. Levallois \emph{et al.}, in preparation.
\bibitem{NovoselovScience04} K.S. Novoselov, A. K.Geim, S. V. Morozov, D. Jiang, Y. Zhang, S.V. Dubonos, I.V. Grigorieva, and A.A. Firsov, Science \textbf{306}, 666 (2004).
\bibitem{GaltPR56} J. K. Galt, W. A. Yager, and H. W. Dail, Jr., Phys. Rev. 103, 1586 (1956).
\bibitem{TaftPhilipPR65} E. A. Taft and H. R. Philipp, Phys. Rev. \textbf{138}, A197 (1965).
\bibitem{ToyPRB77} W. W. Toy, M. S. Dresselhaus and G. Dresselhaus, Phys. Rev. B \textbf{15}, 4077 (1977).
\bibitem{LiPRB06} Z. Q. Li, S.-W. Tsai, W. J. Padilla, S. V. Dordevic, K. S. Burch, Y. J. Wang, and D. N. Basov, Phys. Rev. B \textbf{74}, 195404 (2006).
\bibitem{KuzmenkoPRL08} A. B. Kuzmenko, E. van Heumen, F. Carbone, and D. van der Marel, Phys. Rev. Lett. \textbf{100}, 117401 (2008).
\bibitem{LevalloisSSC12} J. Levallois, M. Tran, and A. B. Kuzmenko, Solid State Commun. \textbf{152}, 1294 (2012).
\bibitem{SchneiderThesis} J. Schneider,  {\it PhD thesis}, LNCMI Grenoble (2010).
\bibitem{BergerJPCB04} C. Berger, Z. Song, T. Li, X. Li, A. Y. Ogbazghi, R. Feng, Z. Dai, A. N. Marchenkov, E. H. Conrad, P. N. First, and W. A. de Heer, \emph{J. Phys. Chem. B}, \textbf{108}, 19912 (2004).
\bibitem{SadowskiPRL06} M. L. Sadowski, G. Martinez, M. Potemski, C. Berger, and W. A. de Heer, Phys. Rev. Lett. \textbf{97}, 266405 (2006).
\bibitem{CrasseeNP11} I. Crassee, J. Levallois, A. L.Walter, M. Ostler, A. Bostwick, E. Rotenberg, T. Seyller, D. van der Marel and A. B. Kuzmenko, Nature Physics 7, 48 (2011).
\bibitem{CrasseePRB11} I. Crassee, J. Levallois, D. van der Marel, A. L. Walter, T. Seyller, and A. B. Kuzmenko, Phys. Rev. B \textbf{84}, 035103 (2011).
\bibitem{dosSantosPRL07} J. M. B. Lopes dos Santos, N. M. R. Peres, and A. H. Castro Neto Phys. Rev. Lett. \textbf{99}, 256802 (2007).
\bibitem{OrlitaPotemskiReview} M. Orlita and M. Potemski, Semicond. Sci. Technol. \textbf{25}, 063001 (2010) .
\bibitem{SalghettiSSC99} F. Salghetti-Drioli, P. Wachter and L. Degiorgi, Solid State Commun. \textbf{109}, 773 (1999).
\bibitem{OppeneerReview} P.M. Oppeneer, \emph{Handbook on Magnetic Materials ed. by K.H.J. Buschow}, Elsevier, Amsterdam \textbf{13} 229 (2001).
\bibitem{HancockPRL11} J. N. Hancock, J. L. M. van Mechelen, A. B. Kuzmenko, Dirk van der Marel, C. Brüne, E. G. Novik, G. V. Astakhov, H. Buhmann, and L. Molenkamp
Phys. Rev. Lett. \textbf{107}, 136803 (2011).
\bibitem{vanMechelenPRL11} J.L.M. van Mechelen, D. van der Marel, I. Crassee and T. Kolodiazhnyi, Phys. Rev. Lett. \textbf{106} 217601 (2011) .
\bibitem{NeshatOE12} M. Neshat, and N.P. Armitage, Opt. Express \textbf{20}, 29063 (2012).
\bibitem{UbrigOE13} N. Ubrig, I. Crassee, J. Levallois, I. O. Nedoliuk, F. Fromm, M. Kaiser, T. Seyller, and A. B. Kuzmenko, Opt. Express \textbf{21}, 24736 (2013).
\bibitem{CrasseeThesis} I. Crassee,  {\it PhD thesis}, University of Geneva (2013).

\end{thebibliography}

\end{document}